\documentclass[aip,graphicx]{revtex4-1}
\usepackage{graphicx}

\draft 

\begin{document}


\title{The Role of Ultraviolet Photons in Circumstellar Astrochemistry} 



\author{T J Millar}
\email[]{Tom.Millar@qub.ac.uk}
\affiliation{Astrophysics Research Centre, School of Mathematics and Physics, Queen's University Belfast, Belfast BT7 1NN, UK }


\date{\today}

\begin{abstract}
Stars with masses between 1 and 8 solar masses (M$_\odot$) lose large amounts of material in the form of gas and dust in the late stages of stellar evolution, during their Asymptotic Giant Branch phase.  Such stars supply up to 35\% of the dust in the interstellar medium and thus contribute to the material out of which our solar system formed. In addition, the circumstellar envelopes of these stars are sites of complex, organic chemistry with over 80 molecules detected in them. We show that internal ultraviolet photons, either emitted by the star itself or from a close-in, orbiting companion, can significantly alter the chemistry that occurs in the envelopes particularly if the envelope is clumpy in nature. At least for the cases explored here, we find that the presence of a stellar companion, such as a white dwarf star, the  high flux of UV photons destroys H$_2$O in the inner regions of carbon-rich AGB stars to levels below those observed and produces species such as C$^+$ deep in the envelope in contrast to the expectations of traditional descriptions of circumstellar chemistry.
\end{abstract}


\maketitle 

\section{Introduction}
\label{sec:intro}

Mass loss from stars on the Asymptotic Giant Branch (AGB) phase typically lasts a few times 10$^5$ years. Despite this short astronomical time-scale, this period of stellar evolution is one of intense study for a number of reasons: (i) their atmospheres are a major site of dust formation in the Universe; (ii) through their stellar winds, they provide at least 30\% of the dust found in the interstellar medium; (iii) the measurement of isotopic ratios from molecular lines allows one to probe nucleosynthetic processes deep inside the star; (iv) a simple dynamical environment that allows physical properties of the wind, such as density, temperature, expansion velocities and chemical abundances, to be determined as a function of radial distance through the circumstellar envelope (CSE), and (v) a relatively simple `astrochemical laboratory' in the outer CSE, having spherical symmetry, a large-scale constant-velocity expansion and with well-defined chemical processes that are dependent on radial distance from the star.

These CSEs, particularly those around carbon-rich AGB stars - notably IRC+10216 (also known as CW Leo) - are rich in linear carbon-chain species, such as C$_n$H ($n$ = 1-8), C$_n$O ($n$ = 1-3,5), C$_n$N ($n$ = 1-3,5), C$_n$S ($n$ = 1-3,5), HC$_n$N ($n$ = 1-5,7,9), the anions C$_n$N$^-$ ($n$ = 1,3,5) and C$_n$H$^-$ ($n$ = 4,6,8).  One should note, however, that the linear nature of these species provides a favourable aspect for radio astronomical detection of rotational transitions since it minimises the value of the rotational partition function - the intrinsic energy of rotation is spread over fewer energy levels than is the case for non-linear molecules.

The CSE around IRC+10216 is a particular object of interest since over 80 molecules have been detected in it, indeed many for the very first, and some the only, time. 
In part, this is due to the carbon-rich nature of the wind since the carbon atom bonds easily to itself and other elements, in part to its very high mass-loss 
rate, $\dot{M}$, in excess of 10$^{-5}$ solar masses per year (M$_\odot$ yr$^{-1}$), and in part to its proximity to Earth, around 130 parsecs (pc).  It has a 
huge CO envelope extending to an angular radius of some 200 arcsec, or about 1 light year. At its distance from Earth, 1 arcsec is equivalent to about 
2 $\times$ 10$^{15}$ cm. Given its expansion velocity, $v_e$ of about 15 km s$^{-1}$, material takes about 10$^4$ yr to move from the dust formation zone at a 
few stellar radii (R$_*$) to the interstellar medium. 

Early observations of this source indicated that molecules could be grouped into two categories, one in which they appeared close to the stellar photosphere 
and the other in which they appeared in shells of radii 12-20 arcsec and with widths of 2-4 arcsec \citep{bet81,gol84,joh84,gol87}. These observations were 
shown to be consistent with a model in which parent species, such as C$_2$H$_2$, HCN and CO, were formed in the high-temperature chemistry near the photosphere, 
accelerated to terminal velocity through collisions with dust grains, and photodissociated by interstellar UV photons as they moved outwards into a region of 
high interstellar UV flux.  The formation of radical daughter products leads to a rapid chemistry that builds up more complex molecules before they too are 
photodissociated by the interstellar field as they move outward through the envelope.  The high mass-loss rate of IRC+10216 ensures that densities are high and 
reactive time-scales short, on the order of a few hundred years, in this interaction zone.

The early models describing molecule formation in the outer CSE were all based on this description \citep{hug82,hug84,gla87,nej84,nej87,che93,mil00}.  The detection 
of abundant hot water at a temperature of several hundred degrees, with a fractional abundance of around 10$^{-7}$, in IRC+10216 \citep{dec10} led to a fundamental 
reappraisal of these models since they predicted most oxygen would be locked up in CO and very little would be free to form water in the warm inner CSE. Although a 
number of suggestions were made for the formation of this hot water, that currently favoured is that the CSE is clumpy enough to allow interstellar UV photons to 
penetrate unhindered to the dense inner envelope. Once in this inner region, the photons can dissociate $^{13}$CO and SiO with the released O atoms able to react 
with H$_2$ to form water \citep{agu10}.  The importance of non-Local Thermodynamic Equilibrium (LTE) chemistry in the inner CSE was strengthened further by the 
detection \citep{agu15} of methyl cyanide, CH$_3$CN, on an angular scale of 1-2 arcsec, a molecule which is not made efficiently through LTE chemistry.

\section{Chemical Structure} \label{sec:chemistry}
In the following sub-sections, I describe briefly the dominant chemistry as a function of radial distance in the CSE.

\subsection{The Photosphere}
\label{sec:photosphere}
The photospheres of AGB stars consists of hot ($\sim$ 2000-3000\,K), dense ($n$ = 10$^{12}$-10$^{14}$ cm$^{-3}$) gas in which LTE determines the chemical composition of the gas. After H$_2$, CO is the most stable molecule and essentially takes up all of the available oxygen in an C-rich star and all the available carbon in an O-rich star. The excess abundance of oxygen in the latter results in large abundances of molecules such as H$_2$O, SiO and SO, while the excess of carbon in the former leads to high abundances of C$_2$H$_2$ and HCN.

\citet{sha95} present the results of LTE abundances and condensation sequences of molecules as a function of the C/O ratio. Although these condensation calculations can give some information on the refractory species in which the elements are sequestered as dust grains condense from the gas, they ignore the detailed physics of grain nucleation and growth as well as the physical changes, for example, large-scale pulsations, thermal pulses, dredge-up of material from the stellar core to the surface, that the star itself can undergo.

A much more detailed description of chemical equilibrium in O-rich, C-rich and S-type (where C/O $\sim$ 1) AGB stars that takes into account both gas-phase and condensed-phase species among 34 elements has been presented by \citet{agu20}. Their detailed predictions for molecular abundances within 5 R$_*$, calculated through mimimising the total Gibbs free energy of the mixture, are compared to observational determinations where these are available.  They show excellent agreement for many species, including C$_2$H$_2$, HCN, CS, SiS, SiO, SiC$_2$, HCP, HF and HCl in C-rich stars. For O-rich AGB stars, they find such agreement for H$_2$O, CO$_2$, H$_2$S, SO, SiO and PO.  Many inorganic species are also observed in agreement with predictions in the inner envelope, including cyanides and isocyanides such as MgCN, MgNC, NaCN, KCN and CaNC, halides such as AlF, AlCl, NaCl and KCl, and oxides such as TiO, TiO$_2$ and AlO.

\citet{agu20} also note some significant failures in of the modelling, most importantly for the hydrides H$_2$O, SiH$_4$ and PH$_3$ in C-rich stars and NH$_3$ in both C-rich and O-rich stars. The differences are very large, with calculated abundances falling some five to six orders of magnitude below those observed. Such failings have led to suggestions for the formation of these species through non-LTE chemistry, including shock and surface chemistry as discussed below.  In addition to including simple diatomic and triatomic species, they have also used Density Functional Theory to calculate the thermochemical data for Ti$_x$C$_y$ species up to $x$ = 13 and $y$ = 22 to identify the key condensation nuclei of the titanium carbide particles often found at the centre of extra-solar carbonaceous dust grains.

\subsection{Pulsational Shock Waves}
\label{sec:shocks}
Although LTE chemistry is treated in a `steady-state' fashion, AGB stars undergo convective motions, large-scale pulsations and thermal pulses, all of which can lead to spatial and temporal inhomogeneities in physical conditions and to deviations from LTE abundances as well as to isotopic compositions. The effects of pulsational shocks on the LTE abundances have been studied in detail in a series of papers by Cherchneff and co-workers \citep{che11, che12,gob16}.  These periodic shocks drive large changes in both density and especially in temperature which essentially destroy all LTE molecules and allow a new gas-phase chemistry in the cooling post-shock gas. In addition, these dense cooling flows also allow dust grains to nucleate and grow. The shocked gas follows a ballistic trajectory, it is subject to repeated shocks, rising up and falling back in the atmosphere, thus enhancing the time-scale over which dust can grow over several pulsation cycles before radiation pressure on the grains eventually enables both the gas and the dust to escape to the circumstellar envelope. The composition of this gas can be significantly different from that predicted by the LTE models.

In particular, \citet{che11} has shown that hot water can be produced at an abundance close to that observed in IRC+10216 from O atoms released in the collisional destruction of CO in the hot post-shock gas.

\subsection{Dust Formation}
\label{sec:dust}
The formation of dust grains in AGB stars has traditionally been described by nucleation theory in which grains grow by the addition of monomers to seed particles and assumes chemical equilibrium \citep{gai88, gai99, hel01}. More recently, there have been attempts to describe the kinetic growth of dust grains, or more precisely, grain seeds, through using quantum mechanical calculations to determine the lowest energy structures and pathways to growth. Thus, Goumans and Bromley \cite{gou12} performed DFT calculations to identify the lowest energy pathway from SiO to Mg$_4$Si$_2$O$_9$H$_2$, through the successive additions of either an oxygen or an Mg or Si atom at each intermediate step. Their calculations showed a common problem on forming grains, namely that the first steps in growth are often highly endoergic. Similar approaches have been made by \citet{bro16}, \citet{gob17} and \citet{bou19} who combine DFT calculations with a chemical kinetic theory of growth and a new approach to nucleation. 

\subsection{Gas-Grain Chemistry}
\label{gasgrain}
With the exception of grain formation at very high temperatures, gas-grain chemistry has been neglected in essentially all circumstellar chemical models to date. This neglect is due to the fact that the time-scales for collisions of the gas with the dust is shorter than the expansion time-scale only at radial distances where the grains are warm enough that few species can remain bound to their surfaces.  Observations do indicate, however, that the gas-grain interaction may be significant even in the warm (T $\sim$ 100-300~K) inner regions of the outflow. Thus, infrared observations of ro-vibrational transitions have shown that fully hydrogenated species such as CH$_4$, NH$_3$, C$_2$H$_4$, and SiH$_4$ \citep{bet79,gol84,kea93}, have abundances that appear to increase at radial distances of around 10-40 stellar radii in IRC+10216. In the interstellar medium, large abundances of fully saturated molecules are associated with Hot Molecular Cores, regions of high mass star formation with temperatures in the range 100--300~K and H$_2$ number densities of 10$^6$--10$^8$ cm$^{-3}$. Such high abundances result from the desorption of ice mantles in which these hydrides have formed through hydrogen atom addition to C, N and O atoms on the ice surface. Although there is no evidence that C-rich AGB envelopes have ice-covered mantles, it is possible that the formation of these hydrides could be the result of reactions involving chemisorbed species on warm, bare carbonaceous grains. The availability of chemisorption sites on bare grains may also serve as a trap for gas-phase molecules and radicals and lead to the removal of species from the gas. In particular, molecules involving silicon, such as SiO, SiC, SiS, SiC$_2$, and Si$_2$C, are often seen to have gas-phase abundances that decrease in this region of the CSE \citep{buj89,sah93,gon03,dec10b,cer10,agu12,cer15}.

\citet{vds19} have recently produced the first model of the circumstellar chemistry that 
includes the gas-grain interaction, with applications to both C-rich and O-rich envelopes. Their model 
includes the accretion of gas-phase species on to the dust grains together with surface chemistry involving 
hydrogenation and reactions between atoms and radicals.  The ice mantles that form can be removed through 
thermal desorption, photodesorption and sputtering of ice with abundant gas-phase molecules due to 
differential velocities in the outflow between the gas and the dust particles. The authors find that 
abundances can be significantly altered in some circumstances depending on the choice of model parameters: 
mass-loss rate and expansion velocity -- increasing $\dot{M}/v_e$ increases the density in the CSE and 
thereby shortens the collisional time between gas and grains; grain temperature -- which depends on the 
material composition, size and morphology of the dust particles. Surface chemistry is enhanced by faster 
diffusion rates on warmer grains; and drift velocity -- sputtering rates increase as this increases. 
Although the binding energies of the gas-phase species are taken from laboratory studies of those in water 
ice mantles rather than on bare carbonaceous or silicate materials, these models are an important first step 
in considering the interaction of gas and dust in CSEs.  Subsequently, \citet{vds20a} 
considered the effects of different grain size distributions and performed radiative transfer calculations 
in order to compare with molecular line observations finding that, in general terms, a combination of their gas-grain model with observation may be able to constrain the size distribution of dust grains in CSE envelopes. 

\subsection{The Outer CSE}
\label{sec:outercse}

Chemistry in the outer region of the CSE is dominated by radiation chemistry, that is, by the interaction of 
the outflowing wind with the external UV radiation field produced by the galactic stars in its neighbourhood. 
Since this interaction generates reactive radicals and ions, it is important to treat the photodissociation of
 abundant molecules as accurately as possible.  In recent years, important advances in this regard have been 
 made in accounting for the self-shielding of CO \cite{sab19} and N$_2$. \citep{li14,li16}

Given the significance of photons to determining the abundances and radial distributions of molecules, we 
devote the following Section to the important topic of photon-driven chemistry.

\section{Photochemistry} \label{sec:photochemistry}

\subsection{External (Interstellar) UV Photons}
\label{sec:extphotons}
Many of the molecules detected through micro- and millimeter-wave rotational spectroscopy show clumpy, shell-like distributions on angular sizes of 10-20 arcsec. These distributions have been interpreted, rather successfully, as arising from the photochemistry induced by the destruction of outflowing, or `parent', molecules from the inner CSE by incoming interstellar UV radiation. For stars such as IRC+10216, an extinction at visible wavelengths of A$_V$ = 1 mag \footnote{Astronomers measure extinction of radiation at wavelength $\lambda$ in magnitudes where A$_{\lambda}$ = 1 corresponds to a change in intensity by 100$^{1/5}$, that is a change in extinction by 5 magnitudes corresponds to a change in intensity of 100.} falls in a region of high density, n $\sim$ 10$^5$ cm$^{-3}$ so that radicals and ions produced by UV photons collide and react to build more complex species on short time scales.  \citet{mil00} showed that a photochemical model could explain both the decreasing column densities of the cyanopolyynes (HC$_{2n+1}$N, $n$ = 1--4) and the increasing radii of their peak abundances as their size increased.  Such models, which assume a spherically symmetric CSE at constant mass-loss rate and outflow velocity, were very successful but do not address some of the more recent evidence that the mass-loss rate is not in steady state nor symmetric. For example, Mauron and Huggins \cite{mau06} showed that the CSE of IRC+10216 contains a series of high density shells, while high spatial resolution observations of the cyanopolyynes and other hydrocarbon chains show that the molecular gas in these shells is clumped.

Brown and Millar \cite{bro03} and Cordiner and Millar \cite{cor09} considered a simple model for these density-enhanced shells and 
showed that they has an impact on the chemistry due to both shorter collisional time scales and enhanced extinction against UV photons 
within the shells. The first `clumpy' model of the outflow was developed to explain the detection of hot water in IRC+10216, \citep{agu10}
 mentioned above in Sect.~\ref{sec:intro}. This model essentially allows a few percent of interstellar UV photons to penetrate deep into 
 the molecular CSE and dissociate $^{13}$CO and SiO unimpeded by the 40 mag or so of dust extinction at UV wavelengths, appropriate for a 
 mass-loss rate of 2 $\times$ 10$^{-5}$ M$_\odot$ yr$^{-1}$ at a radius of 2 $\times$ 10$^{15}$ cm. The liberated O atoms then react quickly with H$_2$ at 
 these warm temperatures to form H$_2$O. The radical OH, the photodissociation product of water is also detected in IRC+10216 \citep{for03}. 
 Note that the abundant $^{12}$CO species is not photodissociated since it self-shields very efficiently\cite{sab19}.

Our detailed chemical kinetic models of the chemistry in the CSEs use the codes and reaction rate coefficients \cite{mce13} publicly available from  the UMIST Database for Astrochemistry website \footnote{www.udfa.net}, with updated photodissociation rates calculated using the cross-sections \cite{hea17} provided in the Leiden photochemistry database \footnote{www.strw.leidenuniv.nl/~ewine/photo}.  The initial, or parent, molecular abundances, that is for those molecules which are formed close to the stellar surface, are taken from \citet{agu10} and shown in Table~\ref{tab:initialabun}. In this paper, we shall consider fractional abundances of species $i$ as a function of radial distance $r$, $x_i(r)$ = $n_i(r)/n_{H_2}(r)$, where we note that the H$_2$ molecule is not photodissociated by the incident radiation field, so that $n_{H_2}(r)$ is  proportional to $r^{-2}$ for an envelope expanding at constant velocity $v_e$.  

We can solve for the fractional abundance, $x_i(r)$, of species $i$ as a function of radius. The ODE that describes the evolution of the 
fractional abundance of $i$ due to both the expansion and chemistry can be written as:

\begin{equation}
\frac{\mathrm{d}\,x_i}{\mathrm{d}r} = \frac{1}{v_e} \left[\sum_{j,k}k_{jk}x_j x_k n_{H_2}(r) + \sum_l k_l
x_l - x_i\left[ \sum_m k_{im}x_m n_{H_2}(r) + \sum_n k_n  \right] \right]\qquad 
\end{equation}

The first and third terms on the right hand side are the summation over all two-body reactions leading the 
formation and destruction of species $i$, respectively, while the second and fourth terms represent one-body reactions, 
that is photo-processes and cosmic-ray ionisations, leading to the formation and destruction of $i$, respectively.

We assume that the radial distribution of the gas temperature is given by a power-law of the form $T(r)$ = $T_{star}$ $(r/R_{star})^p$, 
where $T_{star}$ and $R_{star}$ are the stellar temperature and radius, respectively, and $p$ = -0.7.

\begin{table}
 \caption{Parent species and their initial fractional abundances relative to H$_2$.}
 \label{tab:initialabun}
\begin{tabular}{lclc}
\hline
Species & Abund & Species & Abund \\
\hline
He & 1.7 $\times$ 10$^{-1}$ & CO & 8.0 $\times$ 10$^{-4}$ \\
N$_2$ & 4.0 $\times$ 10$^{-5}$ & C$_{2}$H$_2$ & 8.0 $\times$ 10$^{-5}$ \\
HCN & 2.0 $\times$ 10$^{-5}$ & SiS & 1.0 $\times$ 10$^{-6}$ \\
SiO & 1.2 $\times$ 10$^{-7}$ & CS & 5.0 $\times$ 10$^{-7}$ \\
SiC$_2$ & 5.0 $\times$ 10$^{-8}$ & HCP & 2.5 $\times$ 10$^{-8}$ \\
\hline
\end{tabular}
\end{table}

\begin{figure*}[ht]
\includegraphics[scale = 1.0]{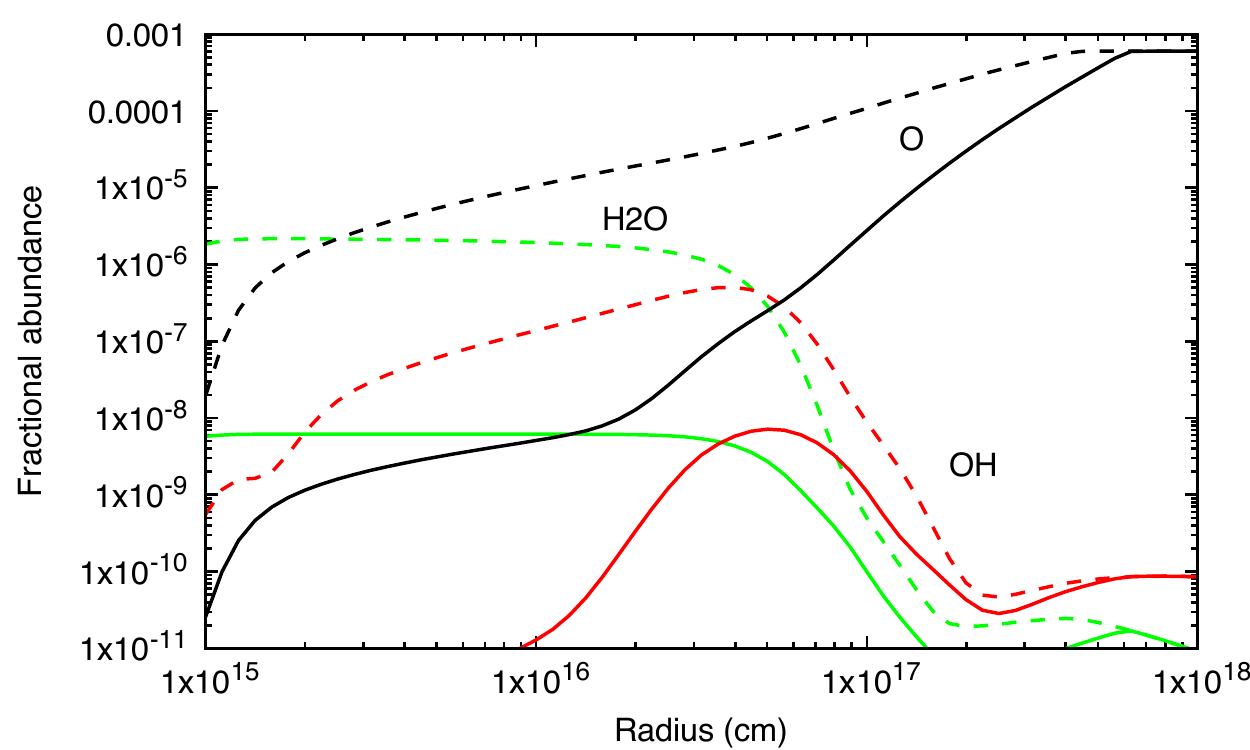}
\caption{\label{fig:water_clumpy} Molecular abundances relative to H$_2$ as a function of radius in a model of IRC+10216 that assumes 2.5\% of interstellar UV photons are able to penetrate to 10$^{15}$ cm without being extinguished by dust grains. The solid curves refer to a `smooth' outflow, that is one in which the density distribution follows a $1/r^2$ distribution. Dashed lines refer to the clumpy model. Colour code: Green, H$_2$O; Red, OH; Black, O.}
\end{figure*}

Fig.~\ref{fig:water_clumpy} shows the radial distribution of water and OH for both a smooth flow and for the `low extinction-type' model 
proposed by \citet{agu10} for parameters typical of IRC+10216, i.e. $\dot{M}$ = 2 $\times$ 10$^{-5}$ M$_\odot$ yr$^{-1}$ and 
$v_e$ = 14.5 km s$^{-1}$. In the latter case I adopt the {\em ad hoc} assumption of \citet{agu10} that 2.5\% of interstellar UV photons 
are not extincted by dust and can 
penetrate down to the innermost envelope with the remaining 97.5\% of photons extinguished by dust. The results confirm those of 
Ag\'{u}ndez et al.  For the standard smooth model, although water can be made in the internal CSE, that is, at radii less than 
10$^{16}$ cm, its abundance is more than two orders of magnitude less than that observed.  The clumpy model on the other hand efficiently 
photodissociates the initial abundance of SiO and converts the oxygen atoms released into H$_2$O at an abundance which agrees with the 
range derived from observation \cite{mel01}, (4-24) $\times$ 10$^{-7}$. Note that the photodissociation rate adopted for SiO at 
10$^{14}$ cm, the start of the outflow, is equivalent to an effective UV extinction of 3.7 in the clumpy medium, much less than the 
its value of about 420 in a smooth outflow. 

This simple model was improved by \citet{vds18a} who investigated the role of clumps and pores on the transfer of interstellar photons to the inner CSE. Instead of an {\it ad hoc} assumption of the fraction of photons able to be transported free of the effects of extinction, they assumed that the CSE is composed of a size distribution of clumps that takes up a fraction $f_{vol}$ of the CSE and which are embedded in an interclump medium characterised by the parameter $f_{ic} = \rho_{ic}/\rho_{sm}$, the ratio of the interclump density to that of a smooth, uniform outflow. Thus, $f_{ic}$ = 1 is equivalent to a smooth outflow, while $f_{ic}$ = 0 represents a void interclump medium, that is a one-component outflow.

\citet{vds18a} calculate an effective UV optical depth for external, interstellar photons, $\tau_{eff}$, from radial distance $r$ to infinity for an exponential distribution of clumps,
\begin{equation}
f(\tau) = \frac{1}{\tau_{cl}}\exp({-\tau/\tau_{cl}})
\end{equation}
where $\tau_{cl}$ is the average optical depth of the clumps. This approach was applied to the chemistries of both O-rich and C-rich CSEs over a range of mass-loss rates from 10$^{-7}$ to 10$^{-5}$ M$_\odot$ yr$^{-1}$. 

\begin{figure*}
\includegraphics[scale = 1.0]{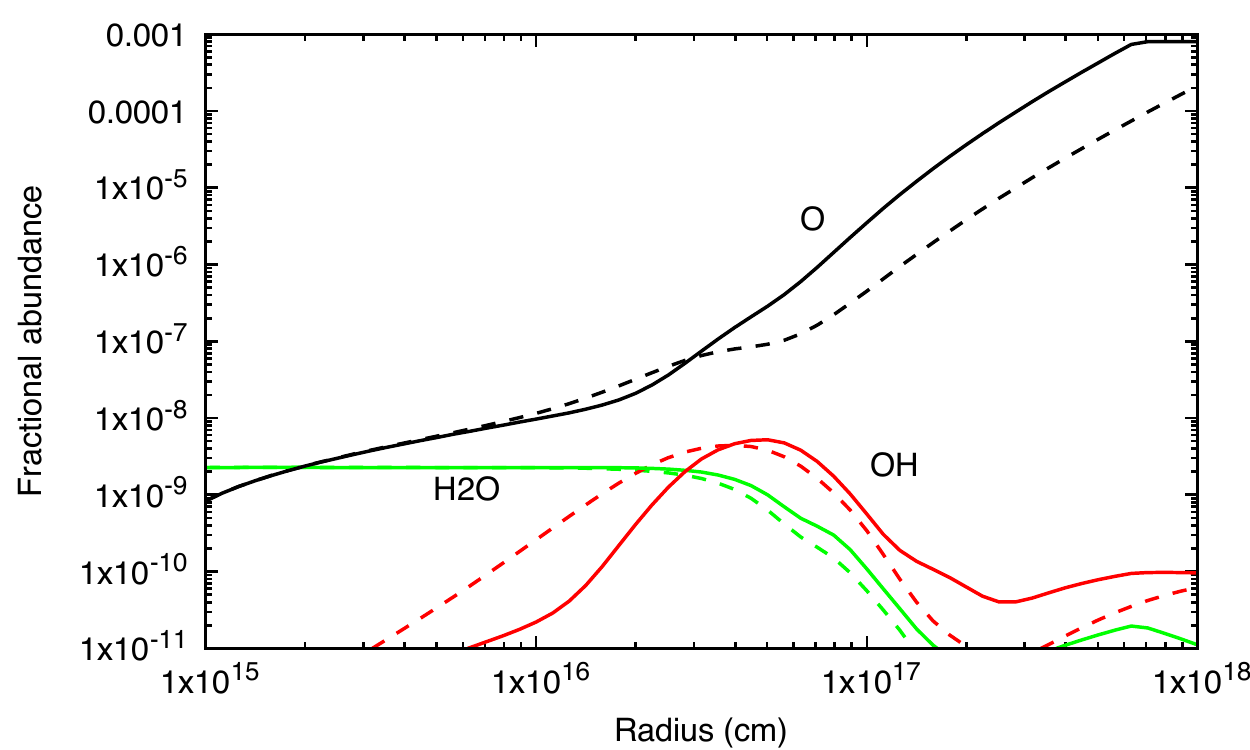}
\caption{\label{fig:water_noIP_0.10.1} As Fig.~\ref{fig:water_clumpy}, except for a porous, clumpy, mass distribution in the CSE. In particular, these results are for a one-component distribution of mass, taking up a fraction 0.1 of the total volume of the CSE. The solid curves refer to the smooth flow, dashed curves to the one-component model. Colour code: Green, H$_2$O; Red, OH; Black, O.}
\end{figure*}

Fig.~\ref{fig:water_noIP_0.10.1} shows the abundance distributions when the approach by \citet{vds18a} is taken for the same physical 
parameters as used in Fig.~\ref{fig:water_clumpy}. The particular case shown here is a one-component model, which minimises the dust extinction seen by 
interstellar photons, with $f_{vol}$ = 0.1, $f_{ic}$ = 0, that is where the mass of the outflow is contained in clumps that take up one-tenth of the total 
volume.  Note that the calculations presented by \citet{vds18a} contained an error in that CO self-shielding was not included - see the correction published by 
\citet{vds20}. This has a large effect on abundances in the inner envelopes of the one-component flows at the highest mass-loss rates, 
10$^{-5}$ M$_{\odot}$ yr$^{-1}$. In particular, the large enhancements derived by \citet{vds18a} for HCN in the O-rich case and 
for H$_2$O in the C-rich case have been over-estimated. Note also that the effects for lower mass-loss rates are less sensitive to this omission since 
column densities of CO are reduced and therefore less sensitive to the effects of self-shielding in these cases. Fig.~\ref{fig:water_noIP_0.10.1} does,
however, include the correct CO self-shielding, although the Van de Sande models do not consider that of N$_2$, an omission that may affect 
their conclusions to some degree, particularly for lower mas-loss rates. One sees here that, for this mass-loss rate, 2 $\times$ 10$^{-5}$ M$_\odot$ yr$^{-1}$, and volume 
filling factor, the radial distribution of the water abundance is essentially identical for both the smooth and clumped outflows and much less than its 
observed value. 

\begin{figure*}
\includegraphics[scale = 1.0]{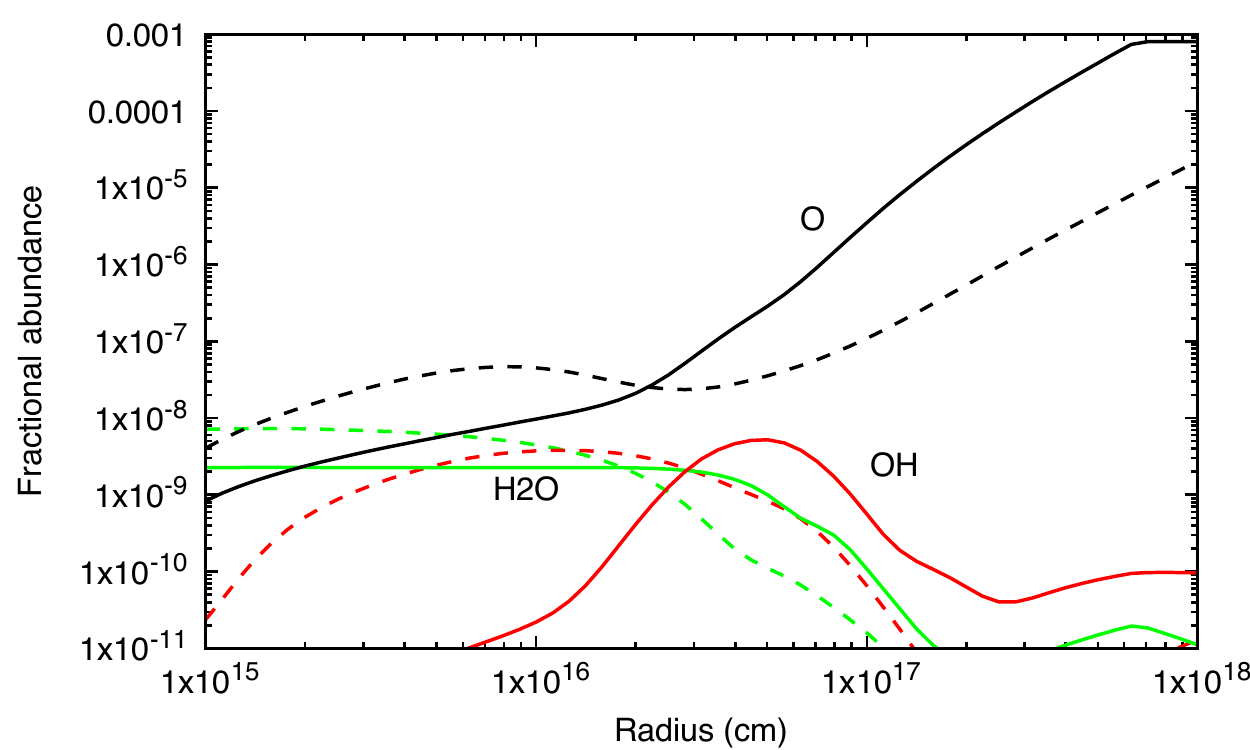}
\caption{\label{fig:water_noIP_0.10.01} As Fig.~\ref{fig:water_noIP_0.10.1}, except that the results are for a one-component distribution of mass, taking up a fraction $f_{vol}$ = 0.01 of the total volume of the CSE. The solid curves refer to the smooth flow, dashed curves to the one-component model. Colour code: Green, H$_2$O; Red, OH; Black, O.}
\end{figure*}

Fig.~\ref{fig:water_noIP_0.10.01} shows, again for a one-component model, that if the outflow is extremely clumpy, $f_{vol}$ = 0.01, then external photons can penetrate to the inner CSE, at a radial distance of less than 10$^{16}$ cm, and increase the water abundance by a factor of a few and the OH abundance by several orders of magnitude. It is noticeable that the radial distribution of OH moves inward in this case due to the increased photodissociation rate of H$_2$O in the inner envelope. However, such small filling factors are inconsistent with the observations of IRC+10216.

\subsection{Internal (Stellar) UV Photons} 
\label{sec:intstellarphotons}
Since AGB stars are cool, with effective temperatures generally less than 3000~K, and form copious amounts of dust, it was rarely 
considered that they might emit sufficient UV photons to affect chemistry. Millar \cite{mil16} was the first to suggest that 
internally generated stellar (blackbody) photons might affect the chemistry of the internal CSE of IRC+10216 ($T_{eff}$ = 2330~K). 
At 50 R$_*$ (or about 1 arcsec at the distance of IRC+10216), the flux of unshielded stellar photons falls to more than 100 times 
less than the interstellar UV flux for $\lambda$ $<$ 1600\AA, so that many species, including CO and N$_2$, are 
unaffected by the presence of these photons.  
Nevertheless, in a clumpy medium, internal stellar photons can play a significant role in driving a {\bf selective} 
chemistry on the spatial scales now being probed by the ALMA interferometer.

For a smooth outflow, the radial extinction in the UV from radius $r$ to infinity is proportional to 1/$r$.  Thus, for the cool AGB stars, 
internal photons have a very large dust extinction as they propagate outwards and play no, or at most a very limited, role in chemistry for 
mass-loss rates above 10$^{-7}$ M$_\odot$ yr$^{-1}$.  If the flows are porous, however, internal photons can find pathways of low extinction 
to the external CSE although their effects are small since, in addition to dust extinction, the photon flux falls off as 1/$r^2$ due to 
geometric dilution. 

\citet{vds19a} applied the porosity formalism to investigate the effects of internal stellar photons on chemistry. They found that for 
mass-loss rates above 10$^{-6}$ M$_\odot$ yr$^{-1}$, the very large effective UV optical depths that occur between the dust formation 
zone and around 10 R$_*$, that is within a radial distance of about 3--5 $\times$ 10$^{14}$ cm, means that there is no effective photochemistry. For 
smaller mass-loss rates, however, smaller effective optical depths mean that these photons can affect chemistry, again in a selective manner,
 and with the largest effects seen in the one-component models which have the lowest optical depths.  Thus, a point at radius $r$ from the 
 star sees external UV photons at an optical depth $\tau_{eff}(r)$ and stellar UV photons at an optical depth, 
\begin{equation}
\Delta\tau_{eff}(r) = \tau_{eff}(r_d) - \tau_{eff}(r)
\end{equation}
where $\tau_{eff}(r_d)$ is the optical depth from the dust formation radius, $r_d$, to infinity.

\begin{figure*}[htb]
\includegraphics[scale = 1.0]{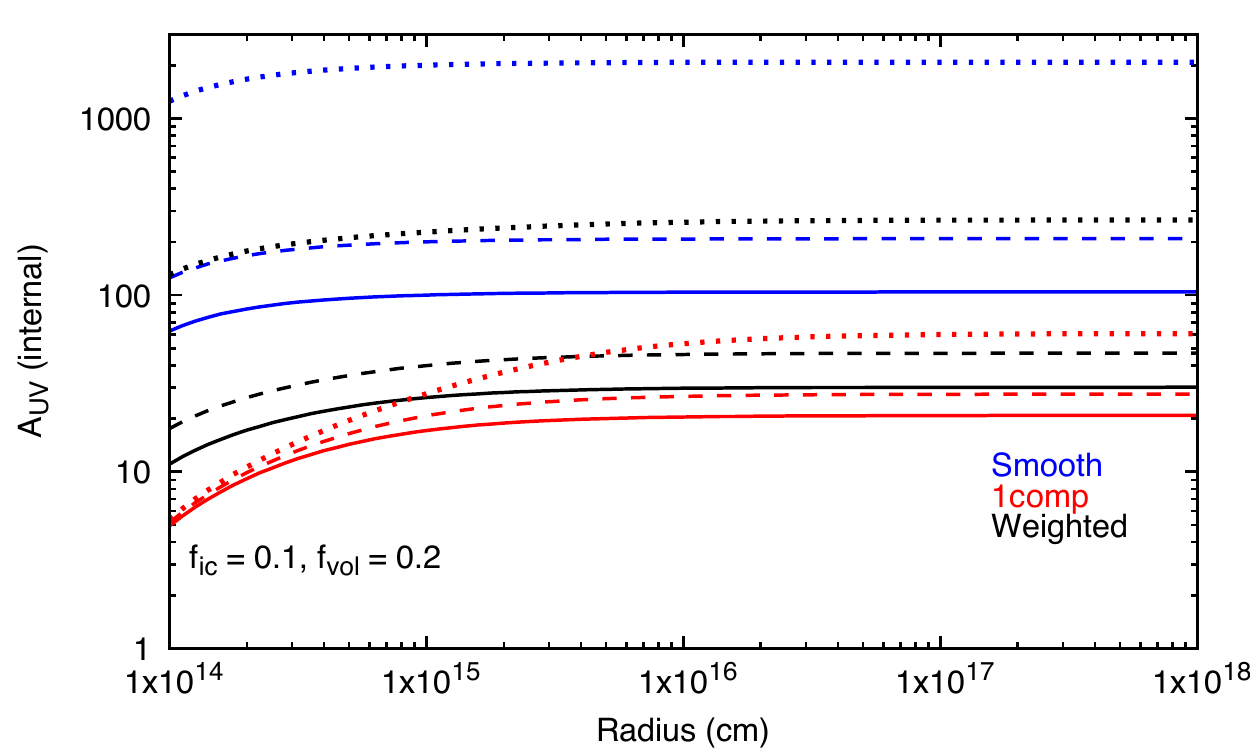}
\caption{\label{fig:auv_eff} Effective radial extinction in magnitudes of internally generated UV photons for mass-loss rates of 10$^{-6}$ M$_\odot$ yr$^{-1}$ (solid lines), 2 $\times$ 10$^{-6}$ M$_\odot$ yr$^{-1}$ (dashed lines), and 2 $\times$ 10$^{-5}$ M$_\odot$ yr$^{-1}$ (dotted lines). The cases of smooth, one-component and two-component (weighted) outflows are color coded.}
\end{figure*}

Fig.~\ref{fig:auv_eff} shows the effective extinction in the UV for internal photons propagating outwards from the star for a variety of 
mass-loss rates for smooth, one-component and two-component (weighted) models. The appropriate rates for photodestruction by internal photons 
are calculated using the cross-sections from \citet{hea17} as:
\begin{equation}
    \beta_i(r) = \beta_i(r_{sc}) \left(\frac{r_{sc}}{r}\right)^2 \exp(-\gamma_i \Delta A_V(r))
\end{equation}
where $\beta_i(r_{sc})$ is the unshielded photorate of species $i$ at a scaling radius $r_{sc}$, and $\Delta A_V(r)$ is the visual extinction from the 
dust formation radius to radial distance $r$.  For those species for which cross-sections are unknown, we use the scaling formula commonly used 
in astrochemical studies of interstellar clouds in which the unshielded interstellar rate, $\beta^{\mathrm{IS}}_0$, is multiplied by the ratio of 
the integrated fluxes of stellar to interstellar photons (integrated over 6--13.6\,eV):

\begin{equation} 
\beta^{\mathrm{IP}}_0  =  \left(\frac{G_*}{G_{\mathrm{IS}}}\right) \beta^{\mathrm{IS}}_0 \label{eqn:integ}
\end{equation}
where 
\begin{equation}
G = \int_{2068\r{A}}^{912\r{A}} F(\lambda) \mathrm{d}\lambda
\end{equation}
and $F(\lambda)$ is the photon flux.

This approach can
significantly overestimate the photoionisation rates for molecules since their ionization potentials generally fall at energies where the 
flux of stellar UV photons depends sensitively on the assumed blackbody temperature of the source. For these species we have
used a reduction factor determined from comparing photoionisation rates calculated for atomic ionisation, for which the exact 
cross-sections are known, to those calculated using the integrated approach.  We include 336 photochannels due to stellar internal 
photons. Our total chemical kinetic network thus includes, when internal photons from a companion object (see next Section) are also included, 
some 6516 reactions among 468 species.

\begin{table}
 \caption{Model parameters. }
 \label{tab:starparam}
\begin{tabular}{llll}
\hline
Species & Abund & Species & Abund \\
\hline
$\dot{M}$ & 2.0 $\times$ 10$^{-5}$ M$_\odot$ yr$^{-1}$ & v$_e$ & 14.5 km s$^{-1}$ \\
T$_{star}$ & 2330~K & R$_{star}$ & 5.0 $\times$ 10$^{13}$ cm \\
R$_{dust}$ & 9.5 $\times$ 10$^{13}$ cm & p  & -0.7 \\
No. Reactions & 6516 & No. Species & 468 \\
\hline
\end{tabular}
\end{table}

\begin{figure*}[htb]
\includegraphics[scale = 1.0]{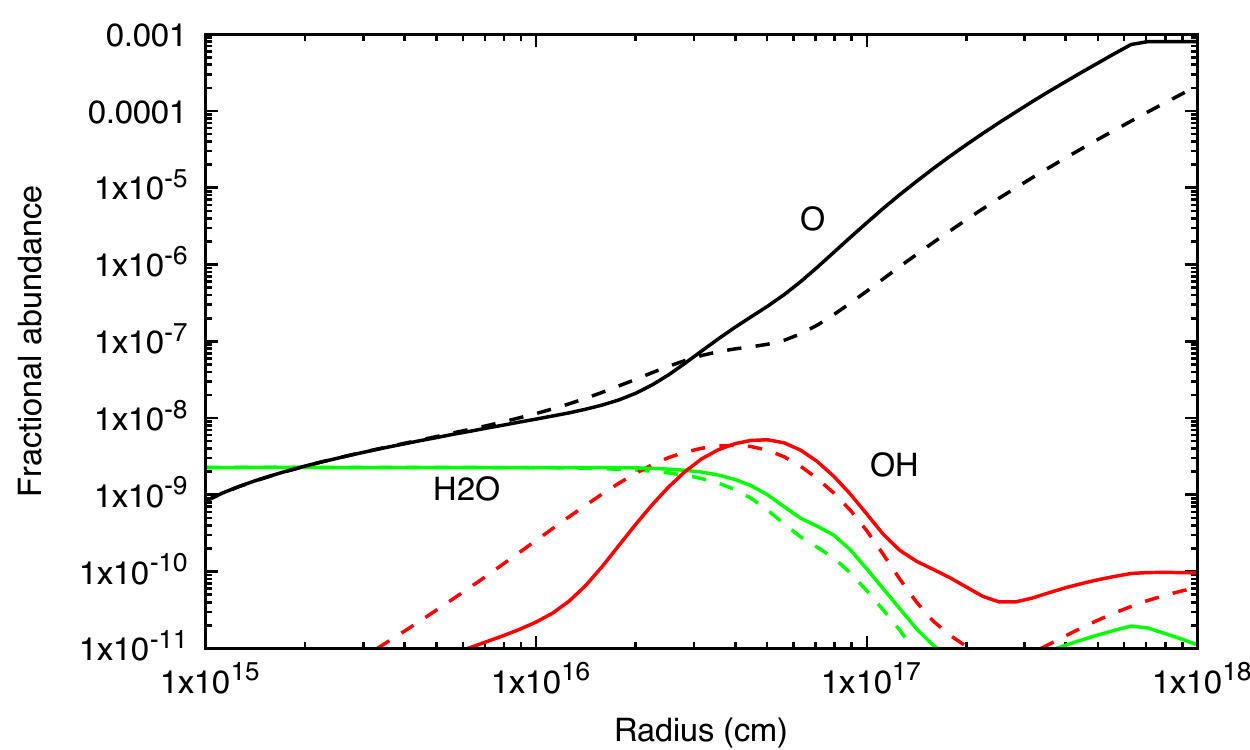}
\caption{\label{fig:water_IP_0101} Water abundance in a model of IRC+10216 that assumes a porous, one component, clump mass distribution in the CSE  with $f_{vol}$ = 0.1 and includes a source of internal UV photons from a stellar blackbody at a temperature of 2330~K. The solid curves refer to the smooth flow, dashed curves to the one-component model. Colour code: Green, H$_2$O; Red, OH; Black, O.}
\end{figure*}

\begin{figure*}
\includegraphics[scale = 1.0]{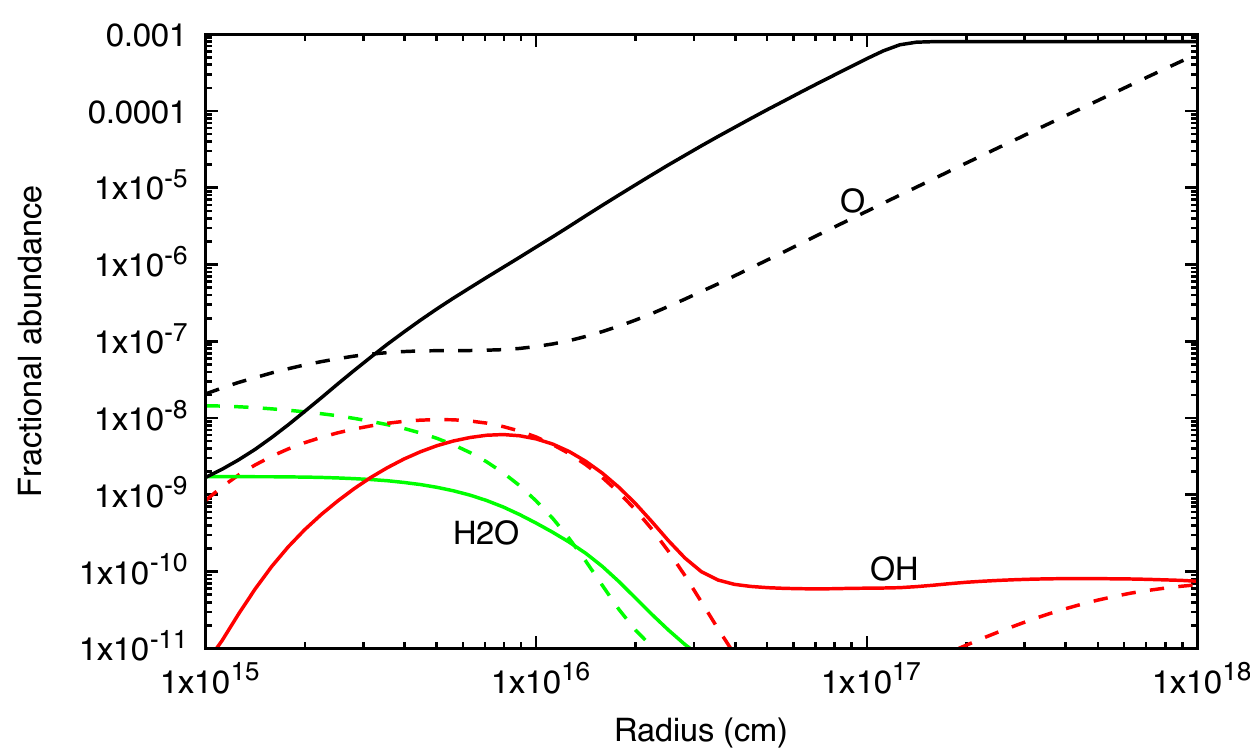}
\caption{\label{fig:water_IP_1.0e-6_01001} As Fig.~\ref{fig:water_noIP_0.10.01}, but for a mass-loss rate of 10$^{-6}$ M$_\odot$ yr$^{-1}$ and a source of internal UV photons from a stellar blackbody at a temperature of 2330~K. Here $f_{vol}$ = 0.01. The solid curves refer to the smooth flow, dashed curves to the one-component model. Colour code: Green, H$_2$O; Red, OH; Black, O.}
\end{figure*}

Table~\ref{tab:starparam} gives the stellar parameters adopted for the models in this Section and  Fig.~\ref{fig:water_IP_0101} shows the results for a model 
similar to that in Fig.~\ref{fig:water_noIP_0.10.1} with the addition of stellar photons at a blackbody temperature of 2330~K, the effective temperature of IRC+10216.  
In this case, we see that the internal, stellar photons do not play any role in increasing the abundance of water. Such low-energy photons can affect molecular 
distributions for smaller mass-loss rates and Fig.~\ref{fig:water_IP_1.0e-6_01001} shows the distributions for $\dot{M}$ = 10$^{-6}$ M$_\odot$ yr$^{-1}$ and an 
extremely clumped outflow with $f_{vol}$ = 0.01. Here one sees an enhancement in the fractional abundance of water by about an order of magnitude at 10$^{15}$ cm.
One should note that the amount of dust and hence extinction is directly proportional to the mass-loss rate. Thus, at radius $r$, both the total and effective 
extinctions are an order of magnitude less in  Fig.~\ref{fig:water_IP_1.0e-6_01001} as those in Figs.~\ref{fig:water_noIP_0.10.1} and \ref{fig:water_IP_0101}.  
As a result, the distributions of water lie closer to the star in the case of lower mass-loss rate since interstellar photons do not suffer as much extinction 
while stellar blackbody photons also suffer less extinction as they propagate outwards through the CSE.

\subsection{Internal (Companion) UV Photons} 
\label{sec:intcompanionphotons}

Stellar blackbody photons may not be the only source of internal UV photons in some AGB stars. Broadband {\em GALEX} observations show that a significant fraction of AGB stars are detectable at UV wavelengths. In a survey of 316 AGB stars, \citet{mon17} find that 179 (57\%) are detected at NUV (2310\AA) and 38 (12\%) detected at FUV (1528\AA) wavelengths. In a careful study, they show that de-reddened, distance-corrected NUV fluxes correlate inversely with $\dot{M}/v_e$, a measure of the H$_2$ number density per unit volume in the CSE, and that in some cases the UV fluxes correlate with the optical light curves, evidence consistent with a stellar, or intrinsic, source for the UV radiation.

In a similar manner, Ortiz and Guerrero \citep{ort16} have shown that main-sequence binary companions of AGB stars can be inferred  from the detection of an AGB star in the GALEX far-UV band and with an observed flux ratio more than 20 times that predicted in the GALEX near-UV band. They show that 34 stars out of a volume-limited sample of 58 fulfill these criteria indicating that they have a main-sequence companion earlier than spectral type K0. Furthermore, they argue that the excess UV emission is not due to a single temperature companion, as might be expected from a star, but may reflect either absorption by the extended atmosphere and CSE of the AGB star or be produced by an accretion flow. Subsequently, \citet{ort19} argued from a survey of some 20 UV-emitting AGB stars that the far-UV, i.e. the highest energy, photons might arise from a hot companion or an accretion disk.

These results indicate that internal UV radiation may be present in a significant fraction of AGB stars, generated either by a stellar companion or by the accretion disk around a star or a planet.

The physical effects of companion objects may also be imprinted on the density structure of the CSE gas. Thus the beautiful spiral patterns detected in IRC+10216 \cite{qui16} and R Sculptoris \cite{mae12}, LL Peg \cite{kim17} and the equatorial density enhancement in L$_2$ Pup \cite{ker14,ker16}, as well as the complex density structures seen in all O-rich AGB stars studied at high spatial resolution in the ALMA Large Programme ATOMIUM \cite{dec20}, can all be interpreted in terms of an orbiting stellar or planetary companion that perturbs the radial outflow from the AGB star.  Recently, \citet{hom20} used ALMA to study the CO and SiO emission in the CSE of the O-rich AGB star EP Aqr at high angular resolution. They found evidence of complex hydrodynamic flows in the inner envelope including a central equatorial density enhancement, which itself contains spiral structures, a bi-conical outflow and a spherical void in the SiO emission at a distance of $\sim$55 au from the star. They argued that the void is due to UV destruction of SiO by a white dwarf companion at a temperature of around 6000~K and a mass between 0.65 and 0.8 M$_\odot$ and, furthermore, that it is this companion that perturbs non-radial motions in the wind of the AGB star. Several theoretical models have been presented showing that the nature of these complex structures depends on the radial momentum in the wind compared to the orbital angular momentum of the companion and therefore on parameters such as the mass-loss rate, the mass ratio of the primary and companion, the orbital radius of the companion, and the stellar wind velocity at the orbital distance of the companion, amongst others.\cite{kim12,che17,elm20}

\begin{figure*}[htb]
\begin{center}
\includegraphics[scale=0.7]{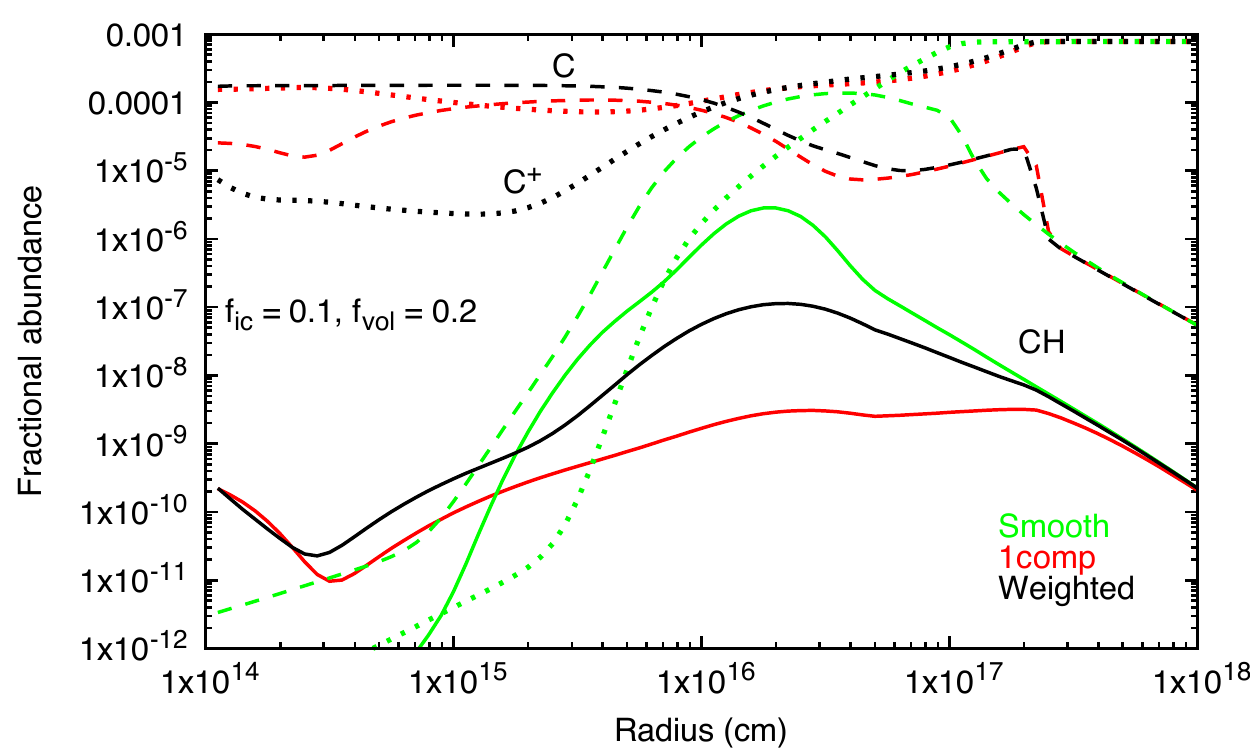}
\includegraphics[scale=0.7]{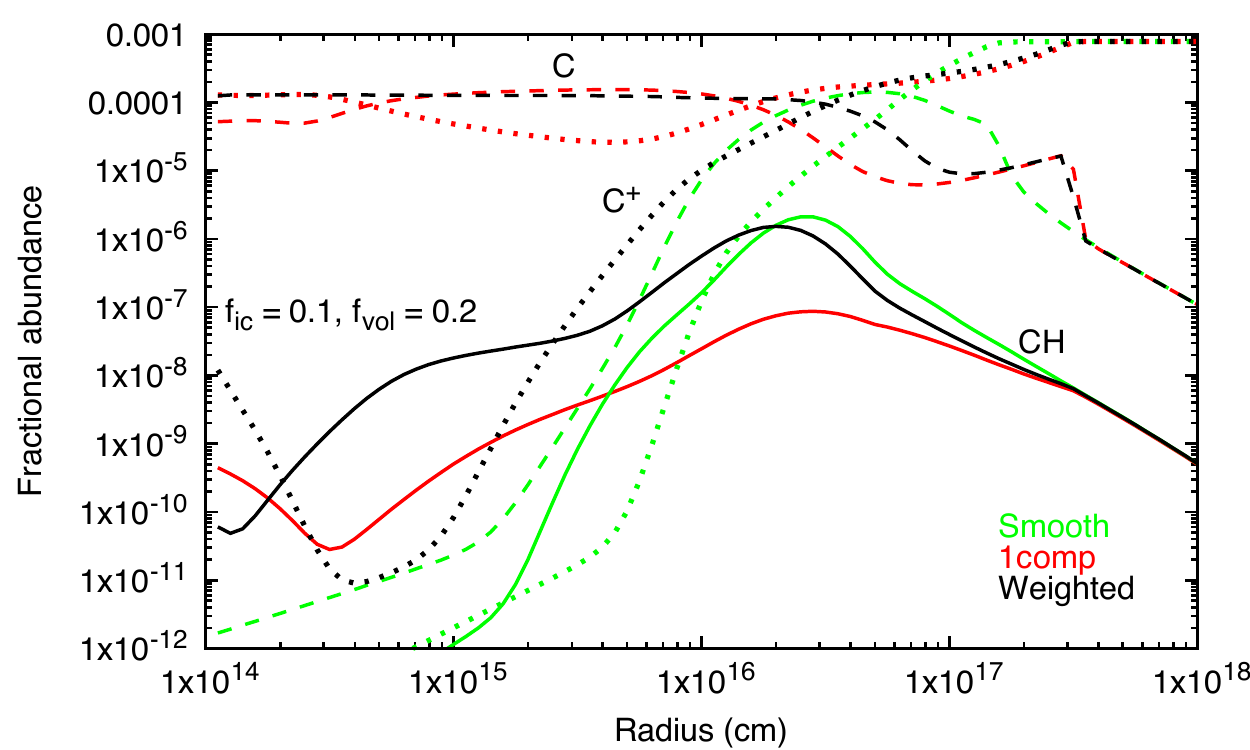}
\includegraphics[scale=0.7]{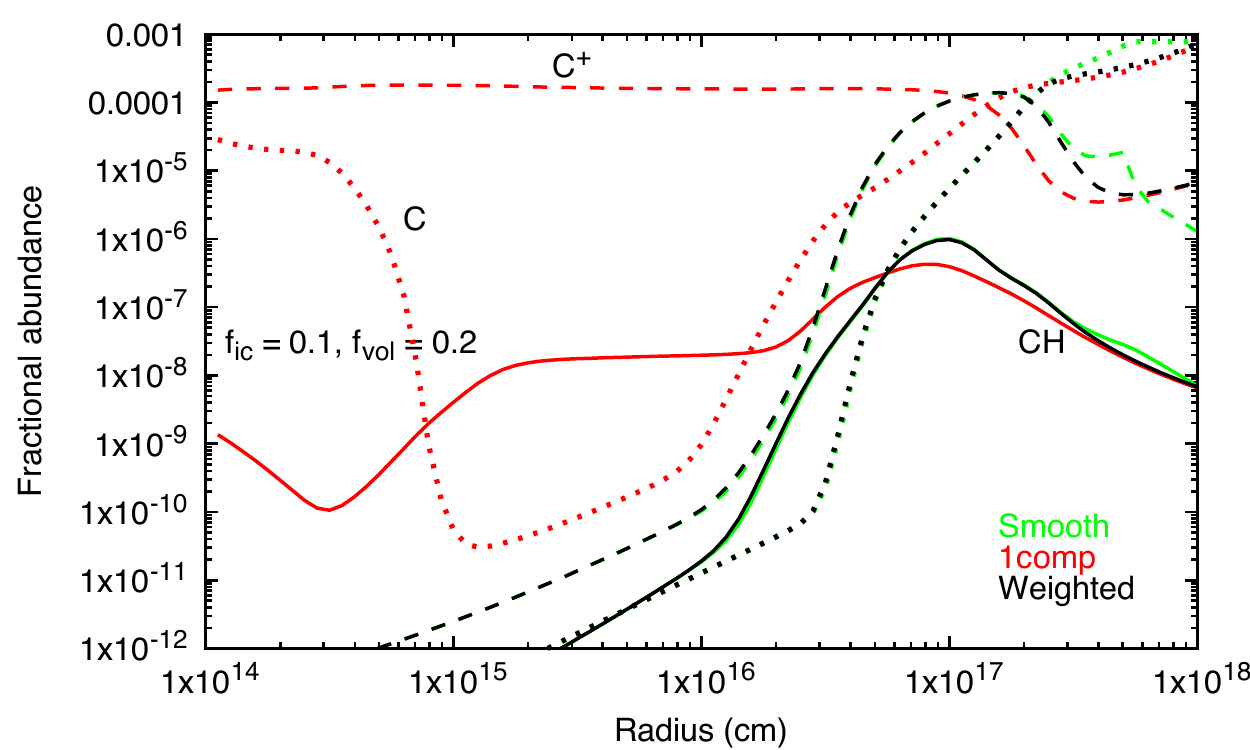}
\end{center}
\caption{Distribution of C, C$^+$ and CH fractional abundances for a stellar BB at 2330~K and a white dwarf companion at 6000~K. Mass-loss rates are 10$^{-6}$ M$_\odot$ yr$^{-1}$ ( top), 2 $\times$ 10$^{-6}$ M$_\odot$ yr$^{-1}$ (middle), and 2 $\times$ 10$^{-5}$ M$_\odot$ yr$^{-1}$ (bottom). Solid lines show the CH abundance, dashed lines that of C, and dotted lines C$^+$.}
\label{fig:C_10-6}
\end{figure*}

In order to examine the effects of UV flux from a companion object, we show in Fig.~\ref{fig:C_10-6} results from a series of representative models 
for a carbon-rich outflow that combine our porosity approach with three sources of UV radiation: (i) the external interstellar flux, (ii) a cool 
(2330~K) blackbody AGB stellar flux, as discussed in Sect.\ref{sec:intstellarphotons}, and (iii) a hotter blackbody flux from a companion object 
such as a white dwarf or an accretion disk. In these models, we choose a white dwarf companion with temperature equal to 6000~K and a radius of 
1.8 $\times$ 10$^{10}$ cm (0.26 R$_{\odot}$). We note that, at this temperature, the neglect of N$_2$ self-shielding does have an 
effect on the abundances calculated for N$_2$ in the inner envelope for the one-component models. In particular, the models may severely underestimate 
the N$_2$ abundance for radii less than a few times 10$^{14}$ cm, although its impact on the abundances of C$^+$ and H$_2$O, discussed below, is likely to be minimal.  In these figures, we show the fractional abundances as a function of radius calculated for three models: 
green curves are for `smooth' outflows, i.e. with density proportional to $r^{-2}$; red curves are for `one-component' models in which the 
interclump density is zero, i.e. $f_{ic}$ = 0; and black curves represent the fractional abundance weighted over those calculated for the 
interclump and clumpy media.  For this we use the formula provided by \citet{vds18a}:

\begin{equation}
    x_{wt}(r) = x_{cl}(r) + f_{ic}(1-f_{vol})(x_{ic}(r) - x_{cl}(r))
\end{equation}
where $x_{wt}, x_{cl}$ and $x_{ic}$ are the weighted, clump and interclump fractional abundances, respectively. These particular models are for the case of $f_{ic}$ = 0.1 and $f_{vol}$ = 0.2 and for mass-loss rates of 10$^{-6}$ M$_\odot$ yr$^{-1}$, 2 $\times$ 10$^{-6}$ M$_\odot$ yr$^{-1}$, and 2 $\times$ 10$^{-5}$ M$_\odot$ yr$^{-1}$.  

The impact of internal, companion UV photons is very significant in the case of low mass-loss rates with the entire CSE showing a significant degree of ionisation 
from the photo-degradation of parent C$_2$H$_2$, HCN and CH$_4$. In these cases it can be seen that, even for mass-loss rates on the order of 
2 $\times$ 10$^{-5}$ M$_\odot$ yr$^{-1}$, abundant C and C$^+$ can be formed just outside the dust formation zone, although the latter's abundance falls 
rapidly as internal photons from the white dwarf get extinguished by dust grains. These species are due to the photodissociation of parent HCN and C$_2$H$_2$
 with C$^+$ most abundant in the highly clumped, one-component outflows with $f_{vol}$ = 0.2 (Fig.~\ref{fig:C_10-6}).  As $f_{vol}$ increases, the outflow 
 becomes `smoother' ($f_{vol}$ = 1) and the ability of internal photons to produce C$^+$ decreases (`smooth' curves, Fig.~\ref{fig:C_10-6}).

\begin{figure*}[htb]
\begin{center}
\includegraphics[scale=0.7]{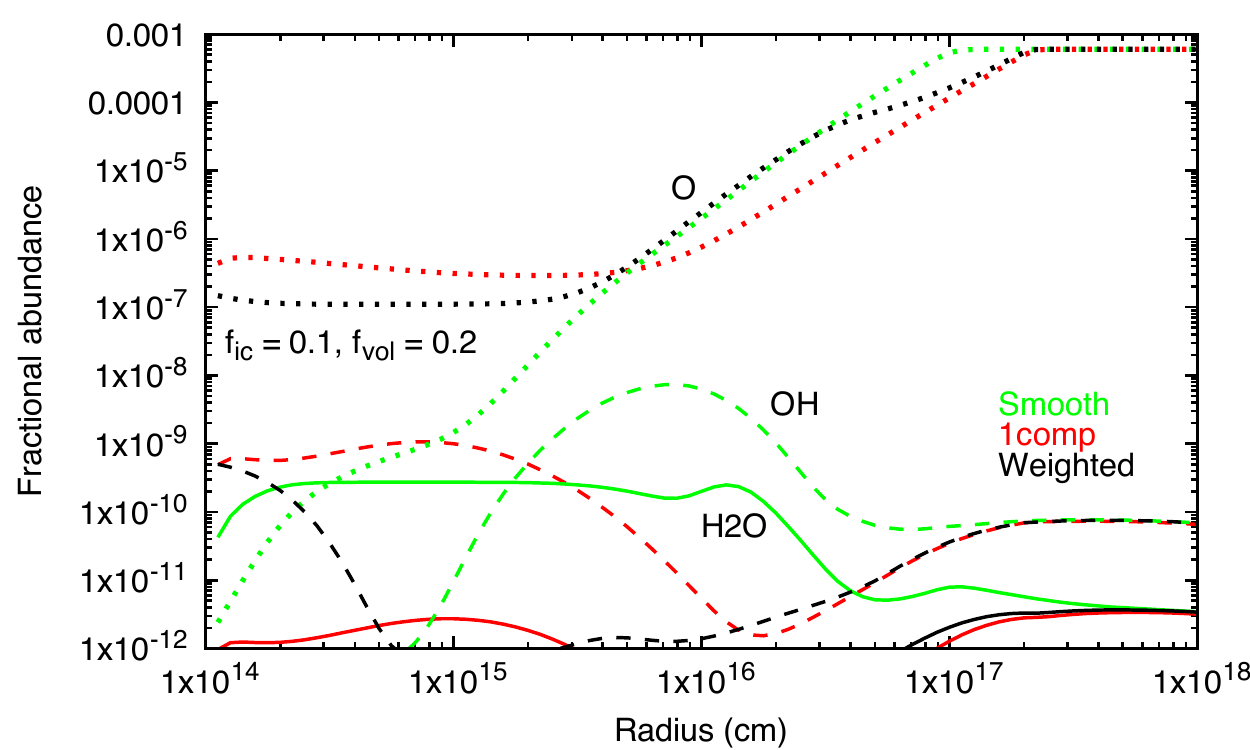}
\includegraphics[scale=0.7]{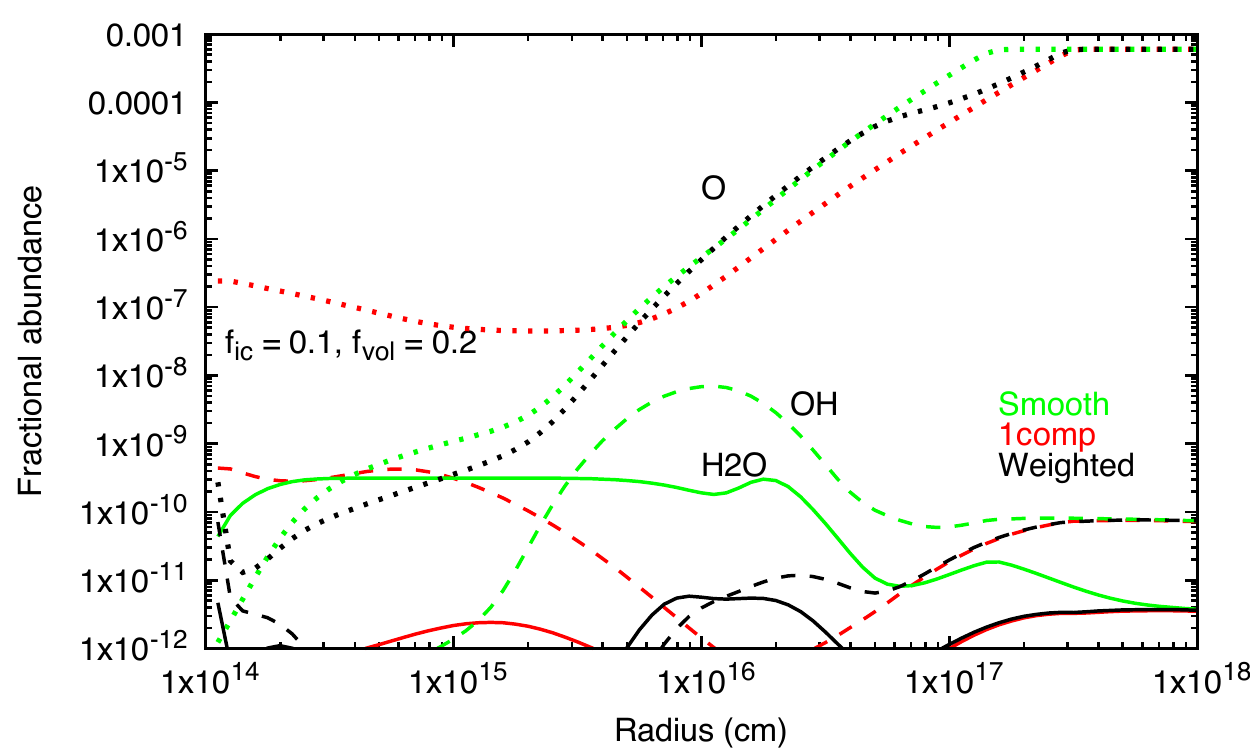}
\includegraphics[scale=0.7]{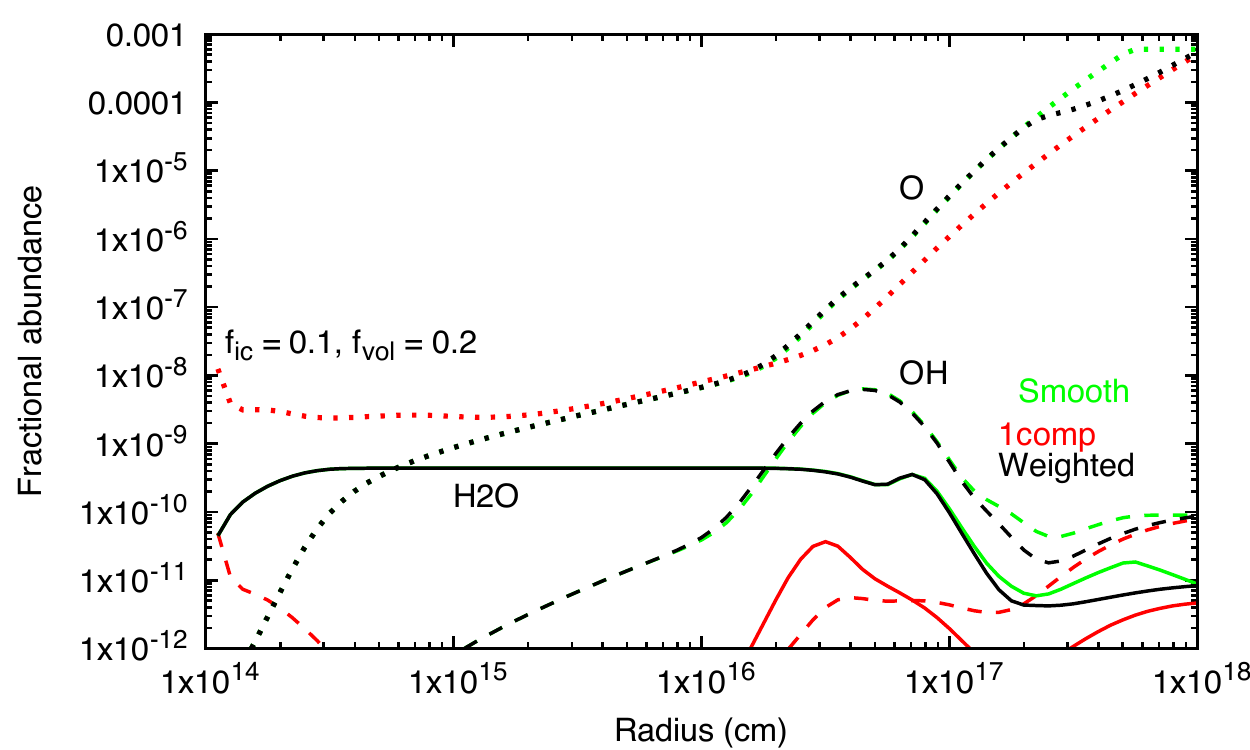}
\end{center}
\caption{Distribution of H$_2$O, OH and O fractional abundances for a stellar BB at 2330~K and a white dwarf companion at 6000~K. Mass-loss rates are 10$^{-6}$ M$_\odot$ yr$^{-1}$ (top), 2 $\times$ 10$^{-6}$ M$_\odot$ yr$^{-1}$ (middle), and 2 $\times$ 10$^{-5}$ M$_\odot$ yr$^{-1}$ (bottom). Solid lines show the H$_2$O abundance, dashed lines that of OH, and dotted lines O.}
\label{fig:H2O_10-6}
\end{figure*}

Finally, we discuss the effect of companion UV photons on the radial distribution of water, which we discussed previously in Sec.~\ref{sec:extphotons} and Sec.~\ref{sec:intstellarphotons}. In both cases we found that the fractional abundance of H$_2$O increased somewhat depending on the value of the mass-loss rate and the degree of porosity with the largest effect seen in the one-component models at low values of $\dot{M}$.
Fig.~\ref{fig:H2O_10-6} shows the distribution of H$_2$O, OH and O for mass-loss rates of 10$^{-6}$ M$_\odot$ yr$^{-1}$, 2 $\times$ 10$^{-6}$ M$_\odot$ yr$^{-1}$, and 2 $\times$ 10$^{-5}$ M$_\odot$ yr$^{-1}$ with an additional companion UV field from a 6000~K blackbody.  

Unlike the cases discussed earlier in Sec.~\ref{sec:intstellarphotons}, in which internal radiation is provided solely by a cool stellar blackbody, the addition of an internal 6000~K radiation field from a white dwarf companion destroys H$_2$O in the inner regions and its abundance is much less, by more than an order of magnitude, than that found when the companion field is absent (see Fig.~\ref{fig:water_IP_0101}) and much less than its observed fractional abundance of $\sim$10$^{-7}$.

\section{Conclusions}
\label{sec:conclusions}

Models of the chemical processes in the circumstellar envelopes of AGB stars that adopt spherical symmetry are still important in describing the observation of molecular emission in the outer envelopes where the majority of molecules are found. It is clear, however, that the density distributions of the inner CSEs, in particular, can no longer be described in terms of spherically symmetric, constant velocity outflows 
irradiated solely by external interstellar UV photons. Furthermore, the fact that AGB stars are cool, with effective temperatures less than 3500~K, typically, means that stellar photons are unable to photodissociate or photoionise many of the common molecules produced near the stellar surface, thereby
limiting their effect on chemistry. Non-smooth outflows are not only driven by stellar phenomena such as thermal pulses or surface 
inhomogeneities but also by the presence of (sub-)stellar companions. As shown conclusively by the ATOMIUM results \cite{dec20}, it is highly likely that the
majority of O-rich AGB stars have a stellar or planetary companion that affects the physical structure of the outflow and which may contribute to the internal 
UV flux that permeates the dust formation zone. While the evidence of companions objects in C-rich objects is less comprehensive, there is no reason to believe 
that the presence of companions will be any less likely in them than those in O-rich stars.  It may be the case, however, that the nature of the density 
perturbations differ in both types of star due to their different dust compositions which can drive larger mass-loss rates in C-rich than in O-rich stars as
carbonaceous grains are more efficient absorbers of stellar radiation than the silicates present in O-rich CSEs. Thus the ratio of radial momentum to orbital 
angular momentum can be larger in C-rich than O-rich stars, making the radially directed winds from the former harder to perturb.

Furthermore, the presence of a non-central, and non-blackbody, UV source, such as a stellar chromosphere, a (sub-)stellar companion, or an accretion disk, means that the transport of photons in a non-symmetric, clumpy medium and its effect on chemistry will be complex to model. The chemical effects of far-UV photons implies that the initial stages of dust formation may involve not only to neutral-neutral reactions, as has been commonly assumed, but also reactions between ions and neutrals. Depending on the nature of the radiation source and the wavelength dependence of the UV flux, chemistry in such regions might be much more selective than is the case in the interstellar medium. That is, only certain species may be dissociated or ionised. For example there may be no photons energetic enough to ionise carbon or sulfur atoms and, as a result, their chemistries may be more restricted than they are in interstellar clouds.

In Sect.~\ref{sec:extphotons}, we showed that external, interstellar UV photons cannot affect the chemistry in the inner CSEs of AGB stars unless either 
the mass-loss rate is low, less than about 10$^{-6}$ M$_\odot$ yr$^{-1}$, or the material is highly clumped so that the effective extinction is around 
two orders of magnitude less than that observed in a smooth outflow.  An alternative means by which UV photons can affect the inner chemistry is to consider 
the role of UV photons generated by the AGB star itself (Sect.~\ref{sec:intstellarphotons})  or by a companion object (Sect.~\ref{sec:intcompanionphotons}).  
In the former case, internal stellar photons are unable to affect the chemistry for mass-loss rates much above 10$^{-6}$ M$_\odot$ yr$^{-1}$ due to the effects of dust extinction \cite{vds19a}. In particular, the large abundance of water detected in IRC+10216 is not reproduced in a model in which $f_{vol}$ = 0.1 due to the large effective extinction in the 
inner envelope. Indeed Fig.~\ref{fig:water_IP_0101} shows that stellar photons do not alter the abundance of H$_2$O at all for radii less than a few times 
10$^{16}$ cm. For a much more porous envelope, for example with $f_{vol}$ = 0.01, then the water abundance increases by about a factor of 3--4 at 10$^{15}$ cm, 
although still at a level about an order of magnitude less than that observed.

If the mass-loss rate is much smaller than that in IRC+10216, for example, 10$^{-6}$ M$_\odot$ yr$^{-1}$, then the water abundance does increase somewhat but its maximum fractional abundance is 10$^{-8}$ at 10$^{15}$ cm, even for a highly porous outflow with $f_{vol}$ = 0.01, though still more than an order of magnitude less than observed (Fig.~\ref{fig:water_IP_1.0e-6_01001}).

The inclusion of an internal, `hot' source of UV photons, for example from a stellar chromosphere, a stellar companion, or an accretion disk around a companion, can change the internal abundances appreciably even for cases in which $f_{vol}$ is relatively large, as shown in Figs.~\ref{fig:C_10-6} and \ref{fig:H2O_10-6}.  The nature of the changes made depends very sensitively on the particular physical parameters chosen, including those associated with the companion -- size, temperature, the wavelength dependence of the UV flux -- and on those associated with the AGB star -- mass-loss rate, wind velocity, clumpiness and porosity of the circumstellar envelope, for example.  These sensitivities will be explored more fully in a future publication.


%
%

%

\begin{acknowledgments}
This manuscript has been improved by the comments of the referees for which I am grateful. I would like to thank the STFC for support under grant number ST/P000312/1 as well as the organisers, in particular Dr Xiaohu Li, of the International Workshop on Astrochemistry, Xi'an, 2019, for their hospitality during my time in Xi'an.
\end{acknowledgments}

\bibliographystyle{apsrev4-1}
\bibliography{xian.bib}

\end{document}